\documentclass[useAMS, usenatbib, usegraphicx]{mn2e}
\usepackage{amssymb}
\usepackage{txfonts}
\usepackage[colorlinks=FALSE, dvips]{hyperref}
\usepackage{journal}

\def\msun{\hbox{${\rm M}_{\odot}$}}
\def\rsun{\hbox{${\rm R}_{\odot}$}}
\def\lsun{\hbox{$L_{\odot}$}}
\def\mstar{\hbox{$M_{\star}$}}
\def\rstar{\hbox{$R_{\star}$}}
\def\lstar{\hbox{$L_{\star}$}}
\def\Prot{\hbox{$P_{\rm rot}$}}
\def\kG{\hbox{$\rm kG$}}
\def\d{\hbox{$\rm d$}}
\def\BI{\hbox{\rm $B_{\rm I}$}}

\def\ie{i.e. }
\def\eg{e.g., }

\begin{document}

\title[Weak and Strong Field Dynamos: from the Earth to the stars]%
  {Weak and Strong Field Dynamos: from the Earth to the stars}

\def\newauthor{%
  \end{author@tabular}\par
  \begin{author@tabular}[t]{@{}l@{}}}
\makeatother
\author[J. Morin et al.] {\vspace{1.7mm}
J.~Morin$^{1}$\thanks{E-mail: jmorin@cp.dias.ie},
E.~Dormy$^2$, M.~Schrinner$^2$,
J.-F.~Donati$^3$\\
$^1$ Dublin Institute for Advanced Studies, School of
Cosmic Physics, 31 Fitzwilliam Place, Dublin 2, Ireland\\
$^2$ MAG (ENS/IPGP), LRA, Ecole Normale Sup\'erieure, 24 Rue Lhomond, 75252 Paris Cedex 05,
France\\
$^3$ IRAP-UMR 5277, CNRS \& Univ de Toulouse, 14 av E Belin, F-31400\\
}
\date{\today,~ Revision:2.01}
\maketitle

\begin{abstract}
Observations of magnetism in very low mass stars recently made important
progress, revealing characteristics that are now to be understood in the
framework of dynamo theory.  In parallel, there is growing evidence that dynamo
processes in these stars share many similarities with planetary dynamos.  We
investigate the extent to which the weak \emph{vs} strong field bistability
predicted for the geodynamo can apply to recent observations of two groups of
very low mass fully-convective stars sharing similar stellar parameters but
generating radically different types of magnetic fields.  Our analysis is based
on previously published spectropolarimetric and spectroscopic data. We argue
that these can be interpreted in the framework of weak and strong field
dynamos.
\end{abstract}


\maketitle

\section{Introduction}
Many stars possessing an outer convective envelope -- like the Sun -- exhibit a
variety of activity phenomena (\eg cool spots producing photometric
variations, a hot corona detected at radio and X-ray wavelengths) powered by
their magnetic field. The latter is generated by dynamo processes converting
kinetic energy (due to thermal convection) into magnetic energy. In the Sun,
and other solar-type stars, the tachocline, a thin shear layer at the base of
the solar convection zone, is thought to play an important role in generating
the magnetic field \cite[\eg][]{Charbonneau97}. On the contrary, main-sequence
stars below $0.35~\msun$ being fully-convective, do not possess a tachocline.
Dynamo processes in these objects are therefore believed to differ
significantly from those in the Sun; in particular, they may operate throughout
the whole stellar interior \cite[\eg][]{Chabrier06, Browning08}.

Measurements of surface magnetic fields on a number of M dwarfs
($0.08<\mstar/\msun<0.7$) have recently been available using two complementary
approaches. One is based on spectroscopy: the average value of the scalar
magnetic field at the surface of the star is inferred from the analysis of the
Zeeman broadening of spectral lines \citep{Saar88, Reiners06}. The other
approach uses time-series of circularly polarised spectra and tomographic
imaging techniques to produce spatially resolved maps of the large-scale
component (typically up to spherical harmonic degree $\ell_{\rm max}$ in the
range 6--30, depending on the rotational velocity) of the vector magnetic field
\cite[Zeeman-Doppler Imaging][]{Semel89,Donati97a,Donati06b}. We refer to
\cite{Morin10b} and references therein for a more detailed comparison. The
first spectropolarimetric observations of a fully-convective M dwarf
\cite[V374~Peg,][]{Donati06, Morin08a} have revealed that these objects can
host magnetic fields featuring a long-lived strong dipolar component almost
aligned with the rotation axis, much more reminiscent of planetary magnetic
fields than those observed on the Sun or solar-type stars
\cite[\eg][]{Donati03}.

Despite many  differences between planetary and stellar interiors, a few recent
studies have strengthened the idea that some fundamental properties of stellar
dynamos could be captured by simple Boussinesq models (i.e. without taking into
account the radial dependency of the stellar density).  An accurate description
of the interior dynamics of stars requires considering their density
stratification. This can be more reliably achieved by anelastic models. It
seems however, that some characteristics are robust enough to be already
captured by a Boussinesq description.  \cite{Goudard09} have shown, for one
given set of parameters, that numerical simulations in the Boussinesq
approximation can reproduce some basic characteristics of {either the magnetism
of planets or partly-convective stars} -- steady axial dipole versus cyclic
dynamo waves respectively -- by varying a single parameter: the relative width
of the convection zone (a thin shell leads to dynamo waves).
\cite{Christensen09} (C09) showed that a scaling law for the magnetic field
strength originally derived from a large number of Boussinesq geodynamo
simulations is also applicable to rapidly-rotating, fully-convective stars
(either main-sequence M dwarfs or young contracting T Tauri stars).

\cite{Featherstone09} show that dynamo action in a fully-convective sphere
(simulating the core of a A-type star) can be strongly enhanced by using a
strong initial dipolar field, as in contrast to the case where only a small
seed field is initially present.  The dipolar solution bears strong
similarities with Boussinesq geodynamo models, as noted by the authors.
Moreover the existence of two co-existing solutions for a given parameter set
seems to be reminiscent of the bistability described in \cite{Simitev09} and
found in Boussinesq models.

Recent spectropolarimetric observations by \cite{Morin10a} (hereafter
M10a) have revealed two radically different types of large-scale magnetism for
M dwarfs with similar masses and rotation periods. One possible explanation for
these observations could be the existence of two dynamo branches in this
parameter regime. In the present letter, we briefly summarize the observational
results on fully-convective stars, describe the theoretical framework of the
weak \emph{vs} strong field dynamo bistability, and discuss its applicability
to very low mass stars.

\section{Large-scale versus total magnetic field}
\label{sec:mdw-bsurf} Following the study of V374~Peg, a first
spectropolarimetric multi-epoch survey was initiated for a sample of 23 active
M dwarfs.  The survey intended to constrain the effects of the shift towards a
fully-convective internal structure on dynamo action \cite[][, M10a]{Morin08b,
Donati08b}{, by mapping the large-scale component of the surface magnetic
field, and assessing the corresponding magnetic flux
\footnote{Here we term ``magnetic flux'' the average modulus of the surface
magnetic field, see \cite{Reiners10b} for a discussion on this term.}
(with a typical uncertainty of the order of 20~\%)}.

All the fully-convective stars of the sample lie in the so called saturated
dynamo regime -- corresponding to $\Prot\lesssim \lbrace5,10\rbrace$~d for a
$\lbrace0.35,0.15\rbrace~\msun$ star \cite[see][]{Kiraga07, Reiners09}.  In
this regime, the rapid dependence of the magnetic flux with the rotation rate
observed for slower rotators (Rossby numbers larger than $0.1$) suddenly stops.
The observations (in terms of X-ray activity or Zeeman broadening) are
consistent with a magnetic field almost independent of the rotation rate.  We
will come back to this point in section~\ref{sec:disc}.  We indeed verified
that for stars for which such measurements exist, the total magnetic
flux inferred from unpolarised spectra is in the range of 1--4~kG (\ie matching
the C09 scaling law).  Two radically different types of large-scale magnetic
fields are observed, either a strong and steady axial dipole field (hereafter
SD) or a weaker multipolar, non-axisymmetric field configuration in rapid
evolution (hereafter WM), whereas no distinction between these two groups of
stars can be made on the basis of mass and rotation only, see
Fig.~\ref{fig:MP}.

\begin{figure}
\includegraphics[width=0.5\textwidth]{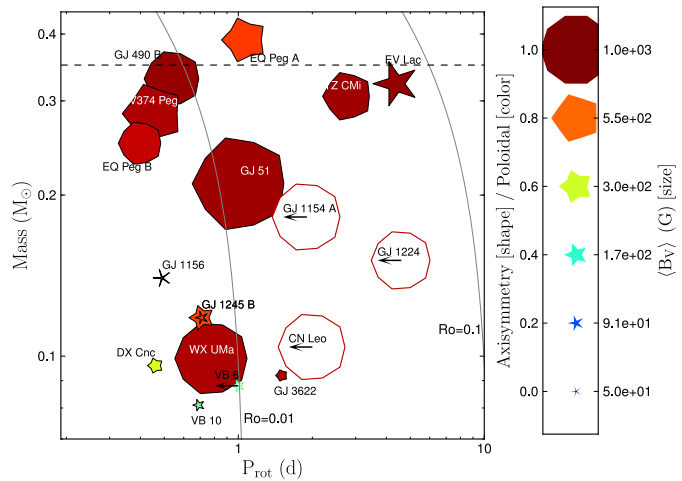}
  \caption[]{Mass--period diagram of fully-convective stars derived from
spectropolarimetric data and Zeeman-Doppler Imaging (ZDI) by
cite{Morin08b}, \cite{Phan-Bao09}, and M10a. Symbol size
represents the reconstructed magnetic energy, the color ranges from blue to red
for purely toroidal to purely poloidal field, and the shape depicts the degree
of axisymmetry from a sharp star for non-axisymmetric to a regular decagon for
axisymmetric. For a few stars of the sample M10a could not perform a
definite ZDI reconstruction,  in these cases only an upper limit of the
rotation period is known and the magnetic flux is extrapolated, those objects
are depicted as empty symbols.  The theoretical fully-convective limit is
depicted as a horizontal dashed line. Thin solid lines represent contours of
constant Rossby number Ro=0.01 (left) and 0.1 (right), as estimated in
M10a.}
  \label{fig:MP}
\end{figure} 

All stars in the strong dipole regime have a typical large-scale magnetic flux
of 1~kG (values comprised between 0.5 and 1.6~kG) whereas for those in the
weak, multipolar regime the typical value is 0.1~kG (all values are lower than
0.2~kG) and it is much more variable for a given object. The latter is only
found in the parameter range $\mstar<0.15~\msun$ and $\Prot<1.5~\d$ in our
sample, though the limits of the domain in which this behaviour occurs are not
yet well defined (see Fig.~\ref{fig:MP},\ref{fig:BV-mdw}), additional
observations on an larger sample of very low mass stars are needed to
specify this point.

\begin{figure*}
  \includegraphics[width=0.50\textwidth]{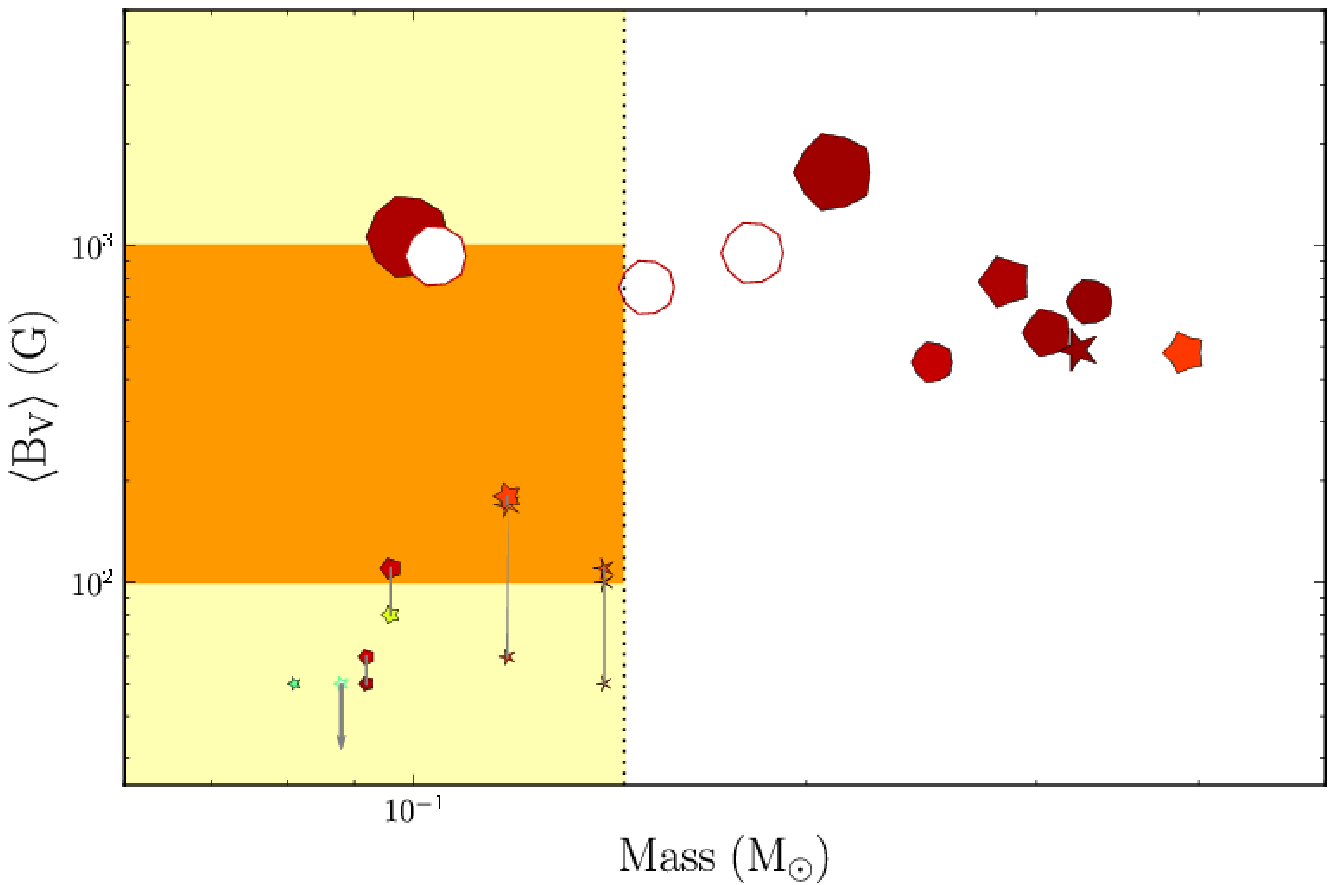}%
  \includegraphics[width=0.50\textwidth]{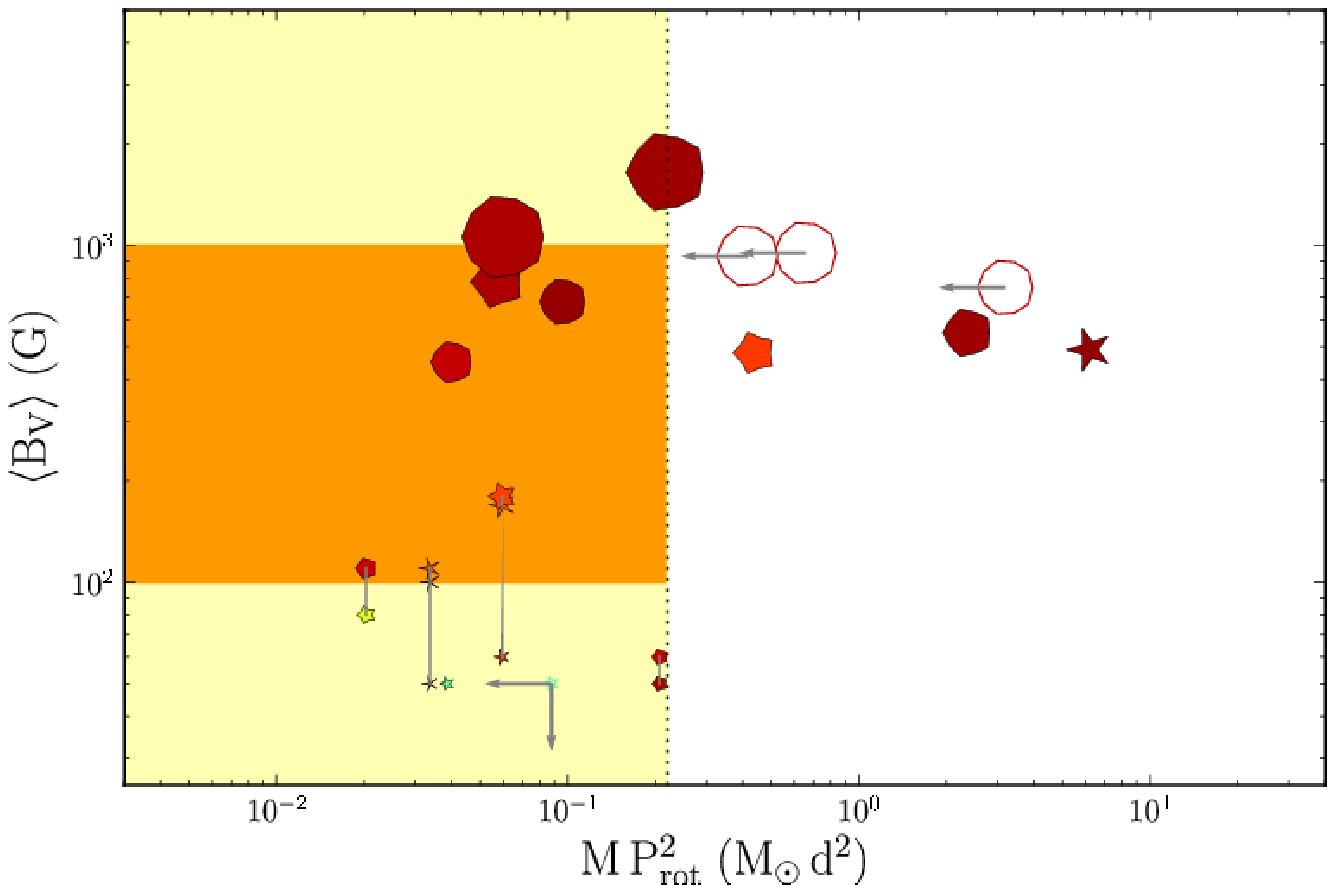}
  \caption[]{Average large-scale magnetic fluxes of fully-convective stars derived from
spectropolarimetric data and Zeeman-Doppler Imaging (ZDI), as a function of
mass (left panel) and mass $\times\,\Prot^2$ (right panel). Symbols are similar
to those used in the mass--period diagram (see Fig.~\ref{fig:MP}). For stars in
the WM regime symbols corresponding to different epochs for a given star are
connected by a vertical grey line. The yellow region represents the domain
where bistability is observed and the orange one separates the two types of
magnetic fields identified (see text).}
  \label{fig:BV-mdw}
\end{figure*}
  
Measurements of the total magnetic flux BI from unpolarised spectroscopy are
available for a number of stars in our spectropolarimetric sample (see
Fig.~\ref{fig:BI-mdw}), with typical uncertainties in the range 0.5--1~kG
\cite[][]{Reiners09, Reiners10}.  The three objects featuring $\BI\sim4~\kG$
are in the SD regime -- large red decagons -- but both SD and WM large-scale
magnetic fields are found among the stars having $\BI\sim 2~\kG$ (see
Fig.~\ref{fig:BI-mdw}). We therefore conclude that there is no systematic
correlation between the unpolarised magnetic flux $\BI$ and the
large-scale magnetic topology inferred from spectropolarimetric observations.
Hence, the two different types of magnetic field configurations are only
detected when considering the large-scale component (probed by
spectropolarimetry, and which represents 15-30~\% of the total flux in the SD
regime, but only a few percent in the WM regime) and not the total magnetic
flux derived from unpolarised spectroscopy.

\begin{figure}
  \includegraphics[width=0.49\textwidth]{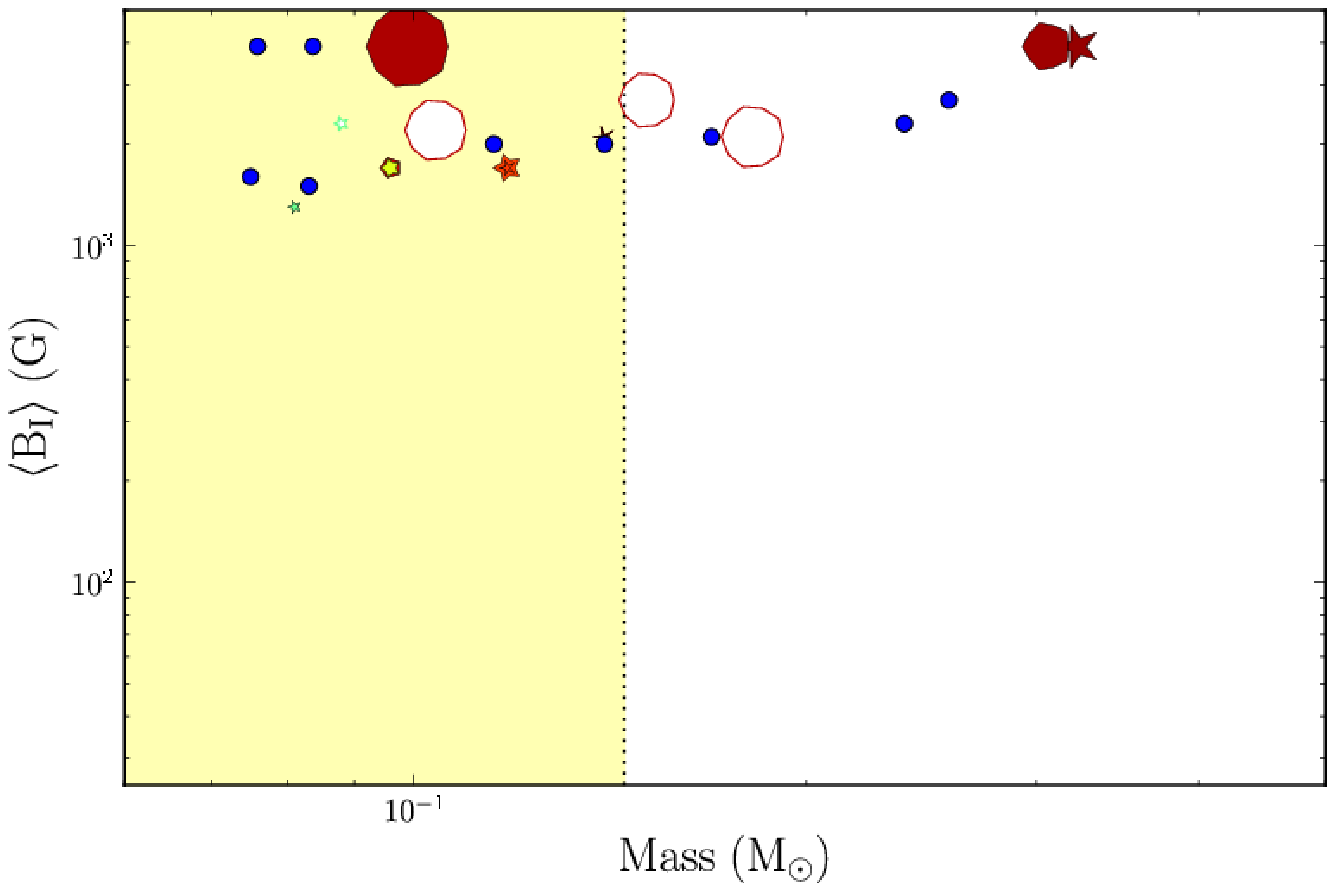}
  \caption[]{Total magnetic fluxes of fully-convective stars in the saturated
  regime measured from unpolarised spectra of FeH lines. The values are taken
  from \cite{Reiners09} and \cite{Reiners10}, whenever 2MASS near infrared
  luminosities \cite[][]{Cutri03} and Hipparcos parallaxes \cite[][]{ESA97} are
  available to compute the stellar mass from the \cite{Delfosse00}
  mass--luminosity relation. Whenever spectropolarimetric data are available
  the properties of the magnetic topology are represented as symbols described
  in Fig.~\ref{fig:MP}. The magnetic field (y-axis) scale is the same as
  Fig.~\ref{fig:BV-mdw}. The yellow region represents the mass domain where
  bistability is observed in spectropolarimetric data (see
  Fig.~\ref{fig:BV-mdw}).  } \label{fig:BI-mdw}
\end{figure}

\section{Weak and strong field dynamos}
\label{sec:wf-sf}
It has been known since \cite{Chandra61} that both, magnetic fields and
rotation, taken separately tend to inhibit convective motions, but that if both
effects are combined the impeding influences of the Lorentz and of the Coriolis
force may be partly relaxed, allowing convection to set in at lower Rayleigh
number and to develop on larger length scales \cite[see also][]{EltayebR70}.
This mutual counteraction of rotation and magnetism is most effective if the
Lorentz and the Coriolis forces are of the same order of magnitude, this is the
{\sl magnetostrophic balance} \cite[see][]{Chandra61, Soward79, Stevenson79}.
This led \cite{Roberts78} to conjecture the existence of two different dynamo
regimes -- a weak and a strong field branch -- and that these different dynamo
solutions could co-exist over some range of parameters \cite[see also][for a
review]{Roberts88}.
\begin{figure}
\centerline{\includegraphics[width=0.35\textwidth]{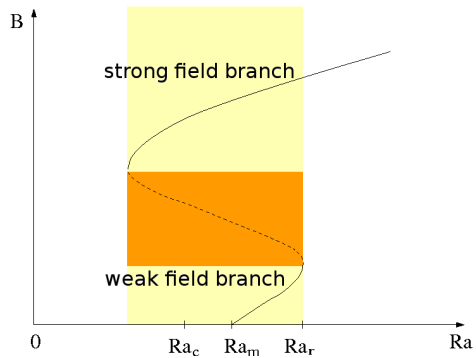}}
\caption[]{Anticipated bifurcation diagram for the geodynamo \cite[adapted
from][]{Roberts88}. The magnetic field amplitude is plotted against the
Rayleigh number. The bifurcation sequence is characterised by two branches,
referred to as weak and strong field branches. The yellow and orange regions
have the same meaning as in Fig.~\ref{fig:BV-mdw}. $\rm Ra_c$ is the
critical Rayleigh number for the onset of non-magnetic convection. The weak
field regime sets in at $ \rm Ra_m$, and the turning point associated with
the runaway growth corresponds to ${\rm Ra} = {\rm Ra_r}$.}
\label{fig:Roberts} 
\end{figure}
The anticipated bifurcation diagram \cite[adapted from][]{Roberts88} is
presented in Fig.~\ref{fig:Roberts}. For dynamos belonging to the weak field
branch, the Proudman-Taylor constraint can only be broken owing to the presence
of the viscous or the inertial term in the momentum equation. This weak field
force balance requires small lengthscales. On the strong field branch, however,
the Lorentz force relaxes the rotational constraint. A similar bifurcation
diagram, but based on the fact that magnetic buoyancy would be negligible close
to the dynamo onset has been proposed for stars by \cite{WT2000}.

The existence of a strong field dynamo regime has received support from
theoretical and numerical studies
\cite[\eg][]{ChildressSoward72,FautrelleChildress82,StPierre93}.  More recently
numerical simulations  have supported the existence of both branches in
spherical geometry, \ie both weak and strong field solutions were obtained
depending on the initial conditions (Dormy \& V.~Morin, \emph{in prep.}).

\section{Discussion}
\label{sec:disc}
We now speculate that the group of stars showing multipolar and time-varying
magnetic topologies (WM) correspond to the weak field regime, whereas those
with a steady dipole (SD) belong to the strong field branch.

The usual control parameter in the weak \emph{vs} strong field dynamo scenario
described above is the Rayleigh number, which measures the energy input
relative to forces opposing the motion. Mass can be used as a good proxy for
the available energy flux in M dwarfs, Fig.~\ref{fig:BV-mdw}a can
therefore be interpreted as a bifurcation diagram for the amplitude of the
large scale magnetic field versus a control parameter measuring the energy
input. In order to compare the driving of convection with the impeding effect
of rotation, we can use $M P_{\rm rot}^2$ as a rough proxy for the
Rayleigh number (see Fig.~\ref{fig:BV-mdw}b) based on rotation rather
than diffusivities \citep[\eg][]{Christensen06}.

Such an identification implies that (i) the strong field/weak field dichotomy
only affects the large-scale component of the magnetic field, (ii) the field
strength is compatible with a Lorentz-inertia force balance for stars featuring
a  WM magnetism, whereas a Lorentz-Coriolis balance prevails for stars in
the SD group. It is difficult to quantify the range of control parameter over 
which both branches co-exist, we will therefore focus our discussion on the 
prevailing force balances in both regimes and their implications on the magnetic 
field.

\subsection{Large-scale dynamo bistability}
Different types of magnetism have previously been found to affect only the
large-scale component of the magnetic field (measured with spectropolarimetry)
and not the total magnetic flux (measured with unpolarised spectroscopy).
Indeed, the aforementioned spectropolarimetric survey has revealed that the
large-scale magnetic field of M dwarfs rapidly changes with stellar mass (both
in geometry and field strength) close to the fully-convective limit
\cite[][]{Morin08b, Donati08b}, whereas no change is visible in total magnetic
flux measurements \cite[][]{Reiners07}.  As the large-scale component only
represents a small fraction of the total flux, a change affecting the
large-scale field alone can indeed remain unnoticed in the measured values of
the total field.

The Rossby number in stars is much higher than in the Earth's interior and
associated with a stronger driving. Therefore, whereas the geodynamo must act
on comparatively large scales (because of the fairly moderate value of the
magnetic Reynolds number) motions in stellar interiors most likely generate
fields on a variety of scales, this includes the possible coexistence of a
large scale dynamo with a small scale dynamo \cite[][]{Voegler_Schuessler2007,
CattaneoH09}. Such co-existence could easily account for the difference in
measurements provided by spectropolarimetry (Fig.~\ref{fig:BV-mdw}) and
unpolarised spectra (Fig.~\ref{fig:BI-mdw}).

\subsection{Force balance and magnetic field strength}
In the strong field regime the Lorentz and Coriolis forces are of the same
order of magnitude. The ratio of the two forces can be estimated by the
Elsasser number:
\begin{equation}
  \Lambda = \frac{B^2}{\rho\mu\eta\Omega},
\end{equation}
where $B$ is the magnetic field strength, $\rho$ the mass density, $\mu$ the
magnetic permeability, $\eta $ the magnetic diffusivity and $\Omega$ the
rotation rate. 

This magnetostrophic force balance is valid for large spatial scales which
are strongly affected by the Coriolis force, and does not apply to small
spatial scales for which the inertial term is predominant in the momentum
equation. It is important to note here that the Elsasser number only provides
a crude measurement for this force balance. To establish this measure, an
equilibrium between induction and diffusion is assumed in the induction
equation. In doing so, the typical length scale of the field and of the flow
have to be considered equal. While this is a sensible approximation for a
planetary dynamo working at moderate magnetic Reynolds number, it is a crude
approximation for stellar interiors. More importantly, this force balance can
only provide an order of magnitude estimate for the field strength.  The
magnetic energy on the strong field branch will obviously vary with the amount
of thermal energy available \cite[see Fig.~\ref{fig:Roberts} and][]{Roberts88}.
Let us nevertheless try to provide an estimate of the surface field
corresponding to an Elsasser number of unity for M dwarfs. We simply take
$\rho=\mstar/(\frac{4}{3}\pi\rstar^3)$, and similarly to C09 we assume that the
ratio between the magnetic field inside the dynamo region and the surface value
is equal to 3.5. An estimate for the turbulent magnetic diffusivity is
also required, crude values, which can be derived from sunspot or active regions 
decay
time or from the formula $\eta\sim u_{rms} \ell$ (where $u_{rms}$ is the
turbulent velocity and $\ell$ is a typical length scale) are {in the range
$10^{11}-3\times10^{12}~{\rm cm^2\,s^{-1}}$ \cite[\eg][]{Ruediger11}}.  Let us 
introduce $\eta_{\rm ref}\equiv 10^{11}~{\rm cm^2\,s^{-1}}$.
With $\eta \propto u_{rms} \ell$, assuming that $u_{rms}$ scales with $\lstar^{1/3}$
according to mixing-length theory \cite[][]{Vitense53}, and $\ell$ with the
depth of the convective zone, we derive an estimate of the field strength at
the stellar surface in the strong field regime:
\begin{equation} \small {B_{sf}} \sim 6\,
\left(\frac{\mstar}{\msun}\right)^{1/2} %
\left(\frac{\rstar}{\rsun}\right)^{-1}  \left(\frac{\lstar}{\lsun}\right)^{1/6}
\mathbf{ \left(\frac{\eta_\odot}{\eta_{\rm ref}}\right)^{1/2}} %
\left(\frac{\Prot}{1~\d}\right)^{{-}1/2}~\kG \end{equation}
Taking stellar radius and luminosity for the stellar mass in the range
$0.08-0.35~\msun$ from \cite{Chabrier97} main sequence models, we note that
$B_{sf}$ is almost independent of mass in this range, and thus the main
dependence is on the rotation period and the chosen reference magnetic
diffusivity $\eta_\odot$. 

However, we do not find evidence for a dependence of the large-scale magnetic
flux on rotation period among stars belonging to the strong field branch in the
spectropolarimetric data. Depending on the precise extent of the bistable
domain, a factor of 2--3 in magnetic fluxes would be expected between the
fastest and slowest rotators of our sample. Such a moderate dependence might
remain undetected due to the dispersion (object-to-object variations) of
measurements.  Using the aforementioned estimate of $\eta_\odot$ and the
rotation periods of the stars in the spectropolarimetric sample, we find
surface values in the strong field regime ranging from 2 to {50}~\kG.  Such
estimates are compatible with the order of magnitude of the measured
large-scale magnetic fluxes. It should be stressed however that this is not
conclusive as the weak field branch is only a factor of ten smaller in
magnitude.

The weak field {\it vs} strong field scenario can however receive
additional support by considering the ratio of field strengths on the
weak and strong field regimes (i.e. the amplitude of the gap between both
branches).  In the case of the Earth's core, inertia and viscous terms become
significant at a similar length scale (because $Ro ^2 E^{-1} \sim {\cal
O}(1)$).  In stellar interiors, however, the Rossby number is usually much
larger than in the Earth's core (by a factor of at least $10^4$).  A weak field
branch would therefore naturally result from a balance between Lorentz and
inertial forces through the Reynolds stresses (i.e. the field strength on the
weak field branch should approach equipartition between kinetic and magnetic
energies).  Therefore, the ratio of field strengths on the weak and strong
field regimes -- corresponding to a different force balance -- is expected to
depend on the ratio between inertia and Coriolis forces. We can thus estimate:
\begin{equation} \frac{B_{wf}}{B_{sf}} = Ro^{1/2}, \end{equation}
where $Ro$ is the Rossby number. M10a derived empirical Rossby numbers
of the order of $10^{-2}$ for stars in the bistable domain, implying a ratio
$B_{sf}/B_{wf}$ of the order of 10, which is indeed the typical ratio of
large-scale magnetic fields measured between the WM and SD groups of stars (see
Fig.~\ref{fig:BV-mdw} and sec.~\ref{sec:mdw-bsurf}).

We shall now discuss an apparent caveat of the proposed scenario. As
noted in section~\ref{sec:mdw-bsurf}, all the stars considered in our sample
belong to the so-called {\it saturated regime} \cite[][]{Kiraga07, Reiners09}.
This means that the strong variation of the field strength with $P_{\rm rot}$
observed for lower rotation rate has ``saturated'' and the  amplitude of the
field seems to be independent of the rotation rate of the star.  Hence, there
is an apparent contradiction with the possibility of a strong field branch, on
which the magnetic field depends on rotation rate as $B_{sf}\propto
\Omega^{1/2}$.
The first important point is that the $B_{sf}\propto \Omega^{1/2}$ (derived
from $\Lambda \sim 1$) should apply here to the large scale field alone,
which is only a fraction of the total magnetic field of the stars (between 15
and 30~\%).  If a small scale dynamo operates, it does not need to follow the
same dependency.  Besides, the slope of this flat portion of the
rotation--magnetic field relation (either of the overall magnetic flux based on
unpolarised spectroscopy or of its proxy the relative X-ray luminosity
$L_X/L_{\rm bol}$) is poorly constrained in the fully-convective regime and is
in fact compatible with a $\Omega^{1/2}$ dependence. %
Evidence for such a dependence of the large-scale magnetic field on 
$\Omega^{1/2}$ would strongly support the proposed scenario.

\section{Conclusion}
In the present letter, we compare the bistability predicted for the geodynamo
with the latest results on spectropolarimetric observations of fully convective
main-sequence stars (M10a). We show that the weak \emph{vs} strong
field dynamo bistability is a promising framework to explain the coexistence of
two different types of large-scale magnetism in very low mass stars. The order
of magnitude of the observed magnetic field in stars hosting a strong dipolar
field (SD), and more conclusively the typical ratio of large-scale magnetic fields
measured in the WM and SD groups of stars are compatible with theoretical
expectations. We argue that the weak dependency of the magnetic field on stellar 
rotation predicted for stars in the strong-field regime cannot be ruled out
by existing data and should be further investigated. We do not make any
prediction on the extent of the bistable domain in terms of stellar
parameters mass and rotation period, this issue shall be investigated by
further theoretical work, and by surveys of activity and magnetism in the 
ultracool dwarf regime.

A dynamo bistability offers the possibility of hysteretic behaviour. Hence the
magnetic properties of a given object depend not only on its present stellar
parameters but also on their past evolution. For instance, for young objects
episodes of strong accretion can significantly modify their structure and hence
the convective energy available to sustain dynamo action \cite[][]{Baraffe09};
initial differences in rotation periods of young stars could also play a role.
Because stellar magnetic fields are central in most physical processes that
control the evolution of mass and rotation of young stars \cite[in particular
accretion-ejections processes and star-disc coupling, \eg][]{Bouvier09,
Gregory10}, the confirmation of stellar dynamo bistability could have a huge
impact on our understanding of formation and evolution of low mass stars.

\section*{ACKNOWLEDGEMENTS} It is a pleasure to thank Katia Ferri\`ere for 
helpful discussions and comments
on this manuscript. We acknowledge the constructive comments of an anonymous
referee.

\bibliographystyle{mn2e}

\end{document}